\documentclass[twocolumn,babel,showpacs,preprintnumbers,amsmath,amssymb]{revtex4}
\usepackage{graphicx}% Include figure files
\usepackage{dcolumn}% Align table columns on decimal point
\usepackage{bm}% bold math

\begin{document}

\preprint{To appear in Physical Review A (2006)}

\title{Momentum transfer for momentum transfer-free which-path experiments}

\author{Aur\'{e}lien~Drezet}
\email{aurelien.drezet@uni-graz.at}
\author{Andreas Hohenau}
\author{Joachim R.~Krenn}
\affiliation{Institute of Physics, Karl-Franzens University,
Universit\"atsplatz 5, A-8010 Graz, Austria}
\date{\today}

\begin{abstract}We analyze the origin of interference disappearance
in which-path double aperture experiments. We show that we can
unambiguously define an observable momentum transfer between the
quantum particle and the path detector and we  prove in particular
that the so called ``momentum transfer free'' experiments can be
in fact logically interpreted in term of momentum transfer.
\end{abstract}

\pacs{03.~65.~Ta, 32.~80.~Lg, 07.~79.~Fc} \maketitle
\section{Introduction}
Bohr's principle of complementarity constitutes the hallmark and
one of the most intriguing features of quantum mechanics. On the
basis of this principle~\cite{Bohr1,Englert} it is indeed
universally accepted that any devices capable of determining the
path taken by a particle trough a Young like double-aperture must
destroy the interference. The justification usually presented is
based on Heisenberg's uncertainty
principle~\cite{Heisenberg,Wootters} and involves an irremediable
exchange of momentum  between the system considered and the
measuring apparatus. Over the last decades the primacy of such
recoil arguments has been however contested in favor of a more
general decoherence mechanism considering the entanglement of the
observed system with its environment~\cite{Zeh}. In particular
Scully \emph{et al.} emphasized~\cite{Scully} that an atom, after
emitting a long wavelength photon in a ``micromaser-cavity'',
located close to one of the two pinholes, can generate a
recordable which-path information without transfer of significant
momentum. The conclusions of~\cite{Scully} seem actually verified
in experiments using entanglement with an internal degree of
freedom to label the path with either neutrons~\cite{Rauch},
photons~\cite{Steinberg,Walborn}, or
atoms~\cite{Eichmann,Rempe}.\\
This stirred-up considerable controversy and a
debate~\cite{Storey,Wiseman,Wiseman1,ScullyR,StoreyR,Bohm,Wiseman2,Scully2,Storey2,Scully3,Yelin}
on the genuine meaning of momentum transfer in which-path
experiments. The paradox comes from the fact that any far-field
interference observed with a Young's like double aperture
experiment is a direct map of the transverse momentum distribution
of the diffracted particle in the aperture plane. Any processes
able to erase the interference should be then interpretable in
term of a perturbation done on this momentum distribution. Such
proposition was formally done~\cite{Storey,Wiseman,Wiseman1} but
the answer is far from being universally
accepted~\cite{ScullyR,StoreyR}. We found two reasons for that:
Firstly, the momentum distribution of the particle in the
single-aperture experiment is not affected by the detector (this
intuitively implies a momentum transfer equal to
zero~\cite{Scully}). Secondly, the momentum transfer defined in
\cite{Storey,Wiseman} is in general ``hidden'' in the sense that
it is not always connected with an experimentally recordable
momentum (like the photon momentum in the Feynman light microscope
\cite{Feynman}).
\\
In this article we revive this controversy by proposing a
consistent definition of momentum transfer based only on the
concept of quantum observables. We study the generality of the
recoil mechanism and show in particular why the scheme proposed in
\cite{Scully} is not recoil-free. We analyze in this context the
roles played by both Heisenberg's relation and entanglement.
\section{Two fundamental examples of which path experiments}
\begin{figure}[h]
\includegraphics[width=3in]{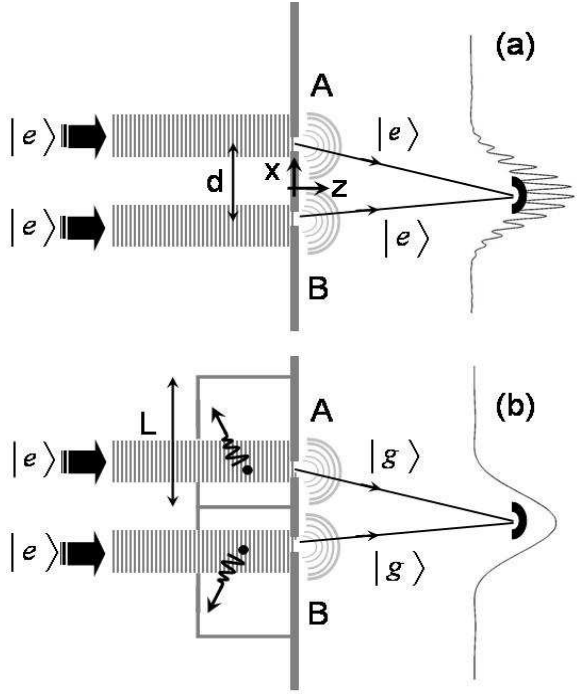}
\caption{Sketch of Young's double-hole experiment for one atom as
discussed in ref.~\cite{Scully} and in the present article. The
separation between the pinholes $A$ and $B$ is $d$ and the
aperture screen is located in the plane $x=0$. (a) Without
which-path detectors we observe interference fringes with high
contrast. (b) Oppositely, if the two micromaser cavities of length
$L\gg d$ are introduced the atom radiates a photon giving the
which path information. As a consequence of Bohr's complementarity
the fringes must disappear.}\label{fig1}
 \end{figure}
We consider at first the scheme suggested in \cite{Scully} and
sketched in Fig.~1. In the absence of any photon emission (see
Fig.~1 (a)) the atomic state immediately behind the holes evolves
into a sum of two diffracted waves:
$\Psi_{I}\left(\mathbf{r}\right)=\Psi_{A}\left(\mathbf{r}\right)+\Psi_{B}\left(\mathbf{r}\right)$,
where $\Psi_{A,B}$ are single aperture wave function reducing to
two narrow peaks located at $\mathbf{r}_{A,B}=\pm
d/2\mathbf{\hat{x}}$ in the aperture plane ($d$ being the distance
between the holes). The which-path detectors consist of two
micromaser empty cavities, one placed at each aperture $A$ and $B$
of a Young interferometer. Before passing through the
double-aperture, an atom initially in a long-lived Rydberg exited
state $|e\rangle$, will radiate a photon in one or the other
cavity and will finish its journey in the ground state $|g\rangle$
(see Fig.~1 (b)). The joint wave function atom-photon is actually
an entangled state which carries the position ambiguity of the
molecule over to an ambiguity of the photon state:
\begin{eqnarray}
|\Psi_{J}\rangle\simeq\int d^{3}\mathbf{r}[
\Psi_{A}\left(\mathbf{r}\right)|\gamma_{A}\rangle+\Psi_{B}\left(\mathbf{r}\right)|\gamma_{B}\rangle]|\mathbf{r}\rangle.
\label{scully}
\end{eqnarray}
Here $|\gamma_{A,B}\rangle$ are single photon states well
localized in one or the other cavity. Such condition can create a
distinguishability since the photon states are orthogonal
e.~g.~$\langle \gamma_{B}|\gamma_{A}\rangle=0$. In the far-field
of the apertures the intensity collected on the screen is
proportional to $G^{(1)}\left(\mathbf{r}\right)=Tr[\hat{\rho}
|\mathbf{r}\rangle\langle\mathbf{r}|]$ (with the density operator
$\hat{\rho}=|\Psi_{J}\rangle\langle\Psi_{J}|$) and we deduce
\begin{eqnarray}
G^{(1)}_{S}\left(\mathbf{r},t\right)\propto
1+\mathcal{V}\cos{[p_{x}d/\hbar+\phi]},\label{decoherent}
\end{eqnarray} where the visibility $\mathcal{V}=|\langle \gamma_{B}|\gamma_{A}\rangle|$ and
phase shift $\phi=\arg{(\langle \gamma_{B}|\gamma_{A}\rangle)}$
are identically equal to zero and where
$p_{x}=\mathbf{\hat{x}}\cdot\mathbf{\hat{r}}h/\lambda_{dB}$ is the
transverse atomic momentum for a particle with de Broglie
wavelength $\lambda_{dB}$. There is clearly a one to one relation
between $G^{(1)}_{S}\left(\mathbf{r},t\right)$ and the momentum
distribution $P(p_{x})$ of the atom in the aperture plane and we
will accept the generality of this relationship in the
following.\\
It is remarkable however that equations similar to
Eqs.~\ref{scully},\ref{decoherent} can be written in the case of
an atom emitting spontaneously a photon while it is still in the
vicinity of the double-pinhole. Supposing the non-relativistic
approximation and that the de-excitation occurs sufficiently fast
behind the pinhole one obtains in this Heisenberg like which path
experiment $\langle
\gamma_{B}|\gamma_{A}\rangle=F_0\left(k_{\gamma}d\right)$
~\cite{Pfau,Chapman,Arndt} which tends to zero for photon
wavelength $\lambda_{\gamma}=2\pi/k_{\gamma}$ much smaller than
$d$. This is indeed in agreement with Heisenberg's back-action
argument.\\
We justify the recoil mechanism at work during the spontaneous
emission process by expanding the photon states in the
wave-vector/polarization-vector basis:
\begin{eqnarray}
|\gamma_{A,B}\rangle=\sum_{\mathbf{k},\mathbf{\epsilon}}\gamma_{\mathbf{k},\mathbf{\epsilon}}^{A,B}|\mathbf{k},
\mathbf{\epsilon}\rangle=\sum_{\mathbf{k},\mathbf{\epsilon}}\gamma_{\mathbf{k},
\mathbf{\epsilon}}^{(0)}e^{-i\mathbf{k}\cdot\mathbf{r}_{A,B}}|\mathbf{k}
,\mathbf{\epsilon}\rangle,\label{photon}
\end{eqnarray}
where $\gamma_{\mathbf{k}, \mathbf{\epsilon}}^{(0)}$ is the photon
amplitude for a hole centered at $x=y=z=0$~\cite{prepreremark}. In
this context Eq.~\ref{scully} becomes
\begin{eqnarray}
|\Psi_{J}\rangle\simeq\int
d^{3}\mathbf{r}\sum_{\mathbf{k},\mathbf{\epsilon}}[
\Psi_{A}\left(\mathbf{r},t\right)\gamma_{\mathbf{k},\mathbf{\epsilon}}^{A}+
\Psi_{B}\left(\mathbf{r},t\right)\gamma_{\mathbf{k},\mathbf{\epsilon}}^{B}]|\mathbf{k},
\mathbf{\epsilon},\mathbf{r}\rangle,\nonumber\\\label{scully2}
\end{eqnarray}
and the atomic intensity recorded on the screen is consequently
\begin{eqnarray}
G^{(1)}_{H}\left(\mathbf{r}\right)\propto
P_{H}(p_{x})=\sum_{\mathbf{k},\mathbf{\epsilon}}P(p_{x},\mathbf{k},\bold{\epsilon}),\label{scully3}
\end{eqnarray} with
\begin{equation}P(p_{x},\mathbf{k},\bold{\epsilon})=
|\tilde{\Psi}^{0}(p_{x}+\hbar
k_{x})|^{2}\cdot|\gamma_{\mathbf{k},\mathbf{\epsilon}}^{(0)}|^{2}[
1+\cos{\left(p_{x}d/\hbar+\mathbf{k}\cdot\mathbf{d}\right)}],
\end{equation}
and where $\tilde{\Psi}^{0}(p_{x})$ is the Fourier transform of
the single aperture wave function $\Psi^{0}(\mathbf{r})$ centered
at the origin. The tiny broadening of the single aperture wave
function is not fundamental and we can make the approximation
$\tilde{\Psi}^{0}(p_{x}+\hbar
k_{x})\simeq\tilde{\Psi}^{0}(p_{x})$ valid for narrow apertures.\\

We can see that the momentum transferred to the photon by the atom
or molecule affects the coherence of the recorded signal. This is
clearly visible from the fact that $G^{(1)}_{H}$ is a sum of
patterns with unit visibility shifted by an amount
$\phi_{\mathbf{k}}=\mathbf{k}\cdot\mathbf{d}=k_{x}d$. Each
individual pattern is unable in itself to erase the fringes but
the sum of all these patterns can do it. Indeed from the values of
the coefficients
$|\gamma_{\mathbf{k},\mathbf{\epsilon}}^{(0)}|^{2}$
\cite{prepreremark,Mandel,ScullyB} we deduce that the uncertainty
$\delta k_{x}$ on the x-component of the photon wave vector is
sufficient to account for the disappearance of fringes (this is
effectively true because $\delta k_{x}\sim
2\pi/\lambda_{\gamma}\gg 2\pi/d$ and $\delta \phi_{\mathbf{k}}\gg
1$). The detailed calculation is straightforward. The coefficients
\begin{equation}
|\gamma_{\mathbf{k},\mathbf{\epsilon}}^{(0)}|^{2}\propto
\frac{|\boldsymbol{\mu}_{ge}\cdot\mathbf{\epsilon}|^{2}
}{\left(\omega-\omega_{\gamma}\right)^{2}+\Gamma^{2}/4},
\end{equation} are obtained for a two level atom with life time
$\Gamma^{-1}$ and transition energy $\hbar\omega_{\gamma}$. After
summing on photon polarization $\epsilon$ one deduces:
\begin{equation}
\sum_{\mathbf{\epsilon}}|\gamma_{\mathbf{k},\mathbf{\epsilon}}^{(0)}|^{2}\propto
\frac{\left(|\boldsymbol{\mu}_{ge}|^{2}-|\boldsymbol{\mu}_{ge}\cdot\mathbf{k}|^{2}/k^{2}\right)
}{\left(\omega-\omega_{\gamma}\right)^{2}+\Gamma^{2}/4}.
\end{equation}
A numerical calculation of Eq.~5 requires evaluation of the sum
$\mathcal{F}(k_{\gamma}d)=\sum_{\mathbf{k}}\frac{\left(|\boldsymbol{\mu}_{ge}|^{2}-|\boldsymbol{\mu}_{ge}\cdot\mathbf{k}|^{2}/k^{2}\right)
}{\left(\omega-\omega_{\gamma}\right)^{2}+\Gamma^{2}/4}e^{i\mathbf{k}\cdot\mathbf{d}}$.
One obtains \cite{prepreremark}
\begin{equation}
\mathcal{F}(k_{\gamma}d)=\big[\sum_{\mathbf{k}}\frac{\left(|\boldsymbol{\mu}_{ge}|^{2}-|\boldsymbol{\mu}_{ge}\cdot\mathbf{k}|^{2}/k^{2}\right)
}{\left(\omega-\omega_{\gamma}\right)^{2}+\Gamma^{2}/4}\big]F_0(k_{\gamma}d)
\end{equation}
and
\begin{eqnarray} G^{(1)}_{H}\left(\mathbf{r},t\right)\propto
1+F_0(k_{\gamma}d)\cos{[p_{x}d/\hbar]},\label{decoherentnew}
\end{eqnarray}
where $F_0(k_{\gamma}d)$ is calculated in
\cite{Paul,Zakowicz,Brukner,Tan}. In particular for an isotropic
distribution of transition dipoles we have
\begin{equation}
\langle F_0(k_{\gamma}d)\rangle=
 \sin{\left(k_{\gamma}d\right)}/\left(k_{\gamma}d\right)e^{-\Gamma
 d/\left(2c\right)}\simeq \sin{\left(k_{\gamma}d\right)}/\left(k_{\gamma}d\right)
\end{equation}
which indeed tends to zero for $\lambda_{\gamma}\ll d$. This
result is in agreement with the complementarity principle since
the fringe visibility depends on the amount of information that is
in principle available to an outside observer. It is consistent
with Heisenberg's back-action argument, since those photons that
provide a better path-information also impart a stronger recoil.
And finally it is derived by decoherence theory, which means it is
based on the entanglement between the atom and the emitted photon.
Equivalently we can write  \begin{equation} P_{H}(p_{x})=
\sum_{\mathbf{k},\mathbf{\epsilon}}|\gamma_{\mathbf{k},\mathbf{\epsilon}}^{(0)}|^{2}P_{0}(p_{x}+\hbar
k_{x}),
\end{equation} with
$P_{0}(p_{x})=|\tilde{\Psi}_{I}(p_{x})|^{2}=|\tilde{\Psi}_{A}(p_{x})+\tilde{\Psi}_{B}(p_{x})|^{2}$.
This is equivalent to Eq.~\ref{scully3} and shows that we can
define a momentum transfer having the effect of a correlation
function. The problem can be then analyzed semiclassically only by
using intuitive arguments based on Heisenberg's relation.\\
At this point it is relevant to repeat that the transfer of
momentum doesn't disturb the single aperture pattern which is not
broadened significantly! This is a central issue here. Indeed in
both the Heisenberg and the Scully \emph{et al.} example the
spatial wave functions $\Psi_{A,B}\left(\mathbf{r},t\right)$
associated with the center of mass are not affected since they
factorize from the photon states. However our treatment of the
Heisenberg which-path experiment considers explicitly the momentum
transfer. This prompts the question as to whether there is or
there is not a recoil mechanism
in the experiment proposed by Scully \emph{et al.}.\\
One should now logically observe that the momentum analysis of
Heisenberg's experiment can be extended to the proposal made in
\cite{Scully}. The photon states in the micromaser can indeed be
expanded like in Eq.~\ref{photon} but with different coefficients
$\gamma_{\mathbf{k}, \mathbf{\epsilon}}^{A,B}$ taking into account
the specifical structure of the photon field confined by the
cavity walls. In the micromaser setup we consider the electric
field being oriented along $x$ and ideally constant inside of the
cavity of width $L$ \cite{Scully,Scully2}. In order to prohibit
overlap between the two cavities we suppose additionally that they
are centered on $x=\pm L/2$. After straightforward calculations we
deduce consequently
\begin{equation}
\gamma_{\mathbf{k}, \mathbf{\epsilon}}^{A,B}\propto
\textrm{sinc}\left(\frac{k_{x}L}{2}\right)e^{\pm
ik_{x}L/2}=\gamma_{k_{x}}^{(0)}e^{\pm ik_{x}L/2},\end{equation}
and we write $G^{(1)}_{S}\left(\mathbf{r}\right)\propto
P_{S}(p_{x})$ with
\begin{eqnarray}
P_{S}(p_{x})=\sum_{k_{x}}|\gamma_{k_{x}}^{(0)}|^{2}[1+\cos{(p_{x}d/\hbar+
k_{x}L)}]|\tilde{\Psi}_{0}(p_{x})|^{2}.\nonumber\\ \label{scully4}
\end{eqnarray}
Again we have a sum of oscillation terms with unit visibility
 like in Eq.~\ref{scully3} but now with a phase shift
$\phi_{\mathbf{k}}=k_{x}L$ instead of $k_{x}d$. Considering this
difference we deduce that in the Scully \emph{et al.} scheme one
only needs a variation of the wave vector in the interval $\delta
k_{x}\sim 2\pi/L\ll 2\pi/d$ in order to destroy the fringes. Such
variation is made possible because the typical dispersion $\delta
k_{x}$ in the Fourier space of the photon is $\sim 2\pi/L$. The
result can be rigorously proven by using the Wiener-Khintchine
theorem. Indeed the total visibility of $G^{(1)}_{S}(\mathbf{r})$
is equivalently defined by
$\mathcal{V}=|\int|\gamma_{k_{x}}^{0}|^{2}e^{ik_{x}L}dk_{x}|$.
From Wiener-Khintchine's theorem we deduce $\int
|\gamma_{k_{x}}^{0}|^{2}e^{ik_{x}L}dk_{x}=\int dx
f(x+L)f^{\ast}(x)=0=\mathcal{V}$ (autocorrelation function) where
$f(x)=\mathcal{F}^{-1}[\gamma_{k_{x}}^{0}]$ is the unit
rectangular function of width $L$ centered on the origin $x=0$.
\\
It is then not true to say that the photon momentum is not
involved in a recoil mechanism since it is only by summing all the
oscillating contribution in Eq.~\ref{scully4} that we can account
for the decoherence effect. \\
The comparison becomes more evident
if we work in the limit of very narrow apertures neglecting the
modification of the single hole diffraction pattern:
\begin{eqnarray}
P_{S}(p_{x})\simeq\sum_{\mathbf{k},\mathbf{\epsilon}}|\gamma_{k_{x}}^{(0)}|^{2}P_{0}(p_{x}+\hbar
k_{x}L/d).
\end{eqnarray}
The approximation $|\tilde{\Psi}_{0}(p_{x}+\hbar
k_{x}L/d)|^{2}\simeq|\tilde{\Psi}_{0}(p_{x})|^{2}$ is here
justified since $\delta k_{x}L/d\sim 1/L\cdot L/d=1/d$ and since
we work in the limit $\delta p_{x}^{\textrm{single hole}}\simeq
h/a\gg h/d$ for the single hole of width $a$. Like in Eq.~12 we
find a correlation function but here the result is not intuitive
because we need an effective momentum  transfer
\begin{equation}
\hbar
k_{x}L/d=\hbar\phi_{\mathbf{k},\bold{\epsilon}}/d\end{equation}
instead of $\hbar k_{x}$. Since
$\hbar\phi_{\mathbf{k},\bold{\epsilon}}/d\gg\hbar k_{x}$ this
proves that the momentum transferred is much higher that the
intuitive and semiclassical expectation $\sim h/L$. This analysis
however does not constitute a disproof but actually a confirmation
of Heisenberg's mechanism in term of recoil transfer because the
effective momentum is typically $\sim h/d$. Both the mathematical
treatment leading to Eqs.~12 and 15 are based on a purely quantum
analysis of momentum transfer. In the case of Eq.~12 the
calculations confirm the intuitive semiclassical reasoning.
However in the case of Eq.~15 only a quantum treatment can justify
the result when the semiclassical analogy fails.
\section{Discussion: which-path information and quantum eraser}
\subsection{Which-path information and momentum transfer}
The precedent result which can seem rather surprising is in fact
very general and not limited to the particular examples of
Eqs.~12,15. Consider the general which-path state given by Eq.~1
but where the detector states are not necessarily associated with
photons and can be expanded in a arbitrary orthonormal basis
$|\xi\rangle$ as
$|\gamma_{A,B}\rangle=\sum_{\xi}\gamma_{\xi}^{A,B}|\xi\rangle$. A
analysis similar to the previous one leads to
\begin{eqnarray}
P(p_{x})=
\sum_{\xi}(|\gamma^{A}_{\xi}|^{2}+|\gamma^{B}_{\xi}|^{2})[
1+\mathcal{V}_{\xi}\nonumber\\
\cdot\cos{\left(\frac{p_{x}d}{\hbar}+\phi_{\xi}\right)}]|\tilde{\Psi}_{0}(p_{x})|^{2},\nonumber\\
\label{scully7bis}
\end{eqnarray}
where $\phi_{\xi} =\arg{\gamma^{A}_{\xi}}-\arg{\gamma^{B}_{\xi}}$
and where $\mathcal{V}_{\xi}$ is the fringes visibility given by
\begin{equation}\mathcal{V}_{\xi}=\frac{
2|\gamma^{A}_{\xi}
|\cdot|\gamma^{B}_{\xi}|}{(|\gamma^{A}_{\xi}|^{2}
+|\gamma^{B}_{\xi}|^{2})}.\end{equation} From the discussion of
the Scully \emph{et al.} and Heisenberg like examples we conclude
that if  (i) $\mathcal{V}_{\xi}$ equals unity for those terms of
the sum Eq.~17 for which
$|\gamma^{A}_{\xi}|^{2}+|\gamma^{B}_{\xi}|^{2}$ has significant
values, and if (ii) in the mean time the phase $\phi_{\xi}$
changes significantly in an interval $\delta\phi_{\xi}\sim 1$ then
decoherence will occur and a momentum transfer can be invoked in
the basis $\xi$ . To be more precise and still in analogy with
Eqs.~12,15
 we can write in the narrow apertures limit \begin{eqnarray} P(p_{x})\simeq
\sum_{\xi}(|\gamma^{A}_{\xi}|^{2}+
|\gamma^{B}_{\xi}|^{2})[(1-\mathcal{V}_{\xi})\nonumber \\
\cdot|\tilde{\Psi}_{0}(p_{x})|^{2}+\mathcal{V}_{\xi}
P_{0}(p_{x}+\hbar\phi_{\xi}/d)] \label{scully7tri}
\end{eqnarray} because we suppose
$\tilde{\Psi}_{0}(p_{x}+\hbar\phi_{\xi}/d)\simeq\tilde{\Psi}_{0}(p_{x})$.
The regime $\mathcal{V}_{\xi}=1$ was the one considered in both
Eqs.~12 and 15. It leads to a simple interpretation in terms of a
correlation function with a momentum transfer
\begin{equation} p_{\xi}=\hbar\phi_{\xi}/d.\end{equation} $\phi_{\xi}$ and then $p_{\xi}$
are clearly experimentally defined by recording the elementary
interference pattern corresponding to $\xi$. In this context
\begin{equation}\delta p_{\xi}\cdot d\simeq \hbar\delta\phi_{\xi}\simeq
\hbar\end{equation} plays the role of Heisenberg's relation. Yet
in general $\mathcal{V}_{\xi}\neq 1$ and this means that there are
some terms in Eq.~19 which are proportional to
$1-\mathcal{V}_{\xi}$ and which cannot be interpreted in terms of
correlation functions and momentum transfers in the basis $\xi$.
At the extreme $\mathcal{V}_{\xi}=0$ and the momentum transfer
discussion is completely irrelevant in the basis $\xi$.\\ However
it is not difficult using the arbitrariness in the basis choice to
find  a representation of the problem in which $V_{\xi}=1$ and in
which momentum transfer is clearly defined. This is true assuming
the ideal which-path experiment
\begin{equation}\langle\gamma_{A}|\gamma_{B}\rangle=0.\end{equation} Indeed
this orthogonality relation means that we can use the two states
$|\gamma_{A,B}\rangle$ as a relevant basis for analyzing the
problem. In this basis $V_{A,B}=0$. However the equivalent choice
\begin{equation}
|\gamma_{\pm}\rangle=(|\gamma_{B}\rangle\pm|\gamma_{A}\rangle)/\sqrt{2}\end{equation}
is completely pertinent too. In such basis $V_{\pm}=1$ and Eq.~20
leads to
\begin{eqnarray}
p_{+}=0,&& p_{-}=\hbar\phi_{-}/d=h/2d.
\end{eqnarray} The
momentum $p_{-}$ accounts for the loss of coherence in the
which-path experiment and we can indeed write
\begin{eqnarray}
P(p_{x})= \sum_{\pm} P_{0}(p_{x}+\hbar\phi_{\pm}/d).
\end{eqnarray}
This means that we can always interpret an ideal which path
experiment in terms of momentum transfer and correlation function
in at least one basis $\xi$.
\subsection{Erasing knowledge with momentum}
The precedent discussion focussed on the concepts of momentum
transfer and which-path information in quantum mechanics. However,
we must add here a few further remarks concerning the quantum
eraser experiment since the existence of a momentum transfer seems
to be in contradiction with the spirit of this experiment. Quantum
erasure which was initially proposed in \cite{Druhl,Scully} and
experimentally realized in
\cite{Zajonc,Steinberg,Herzog,Kim,Walborn} has been recently
discussed in, e.~g., \cite{Mohrhoff,Aharonov}. Such experiments
are usually thought as ``a way around the uncertainty
principle''\cite{Scully,ScullyB,Aharonov} and one says that in
classical physics  ``this question would never come up''
\cite{Scully}. In order to clarify this point one has first to
remind that the philosophy of the quantum eraser proposal is based
on the possibility to rewrite an entangled state like
\begin{equation}
\Psi_A(\mathbf{r},t)|\gamma_{A}\rangle+\Psi_B(\mathbf{r},t)|\gamma_{B}\rangle
\end{equation}
in the equivalent form
\begin{equation}
\Psi_{+}(\mathbf{r},t)|\gamma_{+}\rangle+\Psi_{+}(\mathbf{r},t)|\gamma_{+}\rangle,
\end{equation} with
\begin{equation}
\Psi_{\pm}(\mathbf{r},t)=(\Psi_{A}(\mathbf{r},t)\pm\Psi_B(\mathbf{r},t))/\sqrt{2}
\end{equation}
and where $|\gamma_{\pm}\rangle$ are defined by Eq.~23. \\
Clearly if one is able to project the photon state in the
orthogonal basis $|\gamma_{\pm}\rangle$ one would be able to
retrieves interference fringes (or even anti-interference fringes
if the projection is made on $|\gamma_{-}\rangle$). Since this
projection can be delayed the atom fringes can be rebuild (using
coincidence measurement techniques) even after that the atom
reached the screen \cite{Kim}.\begin{figure}[h]
\includegraphics[width=3in]{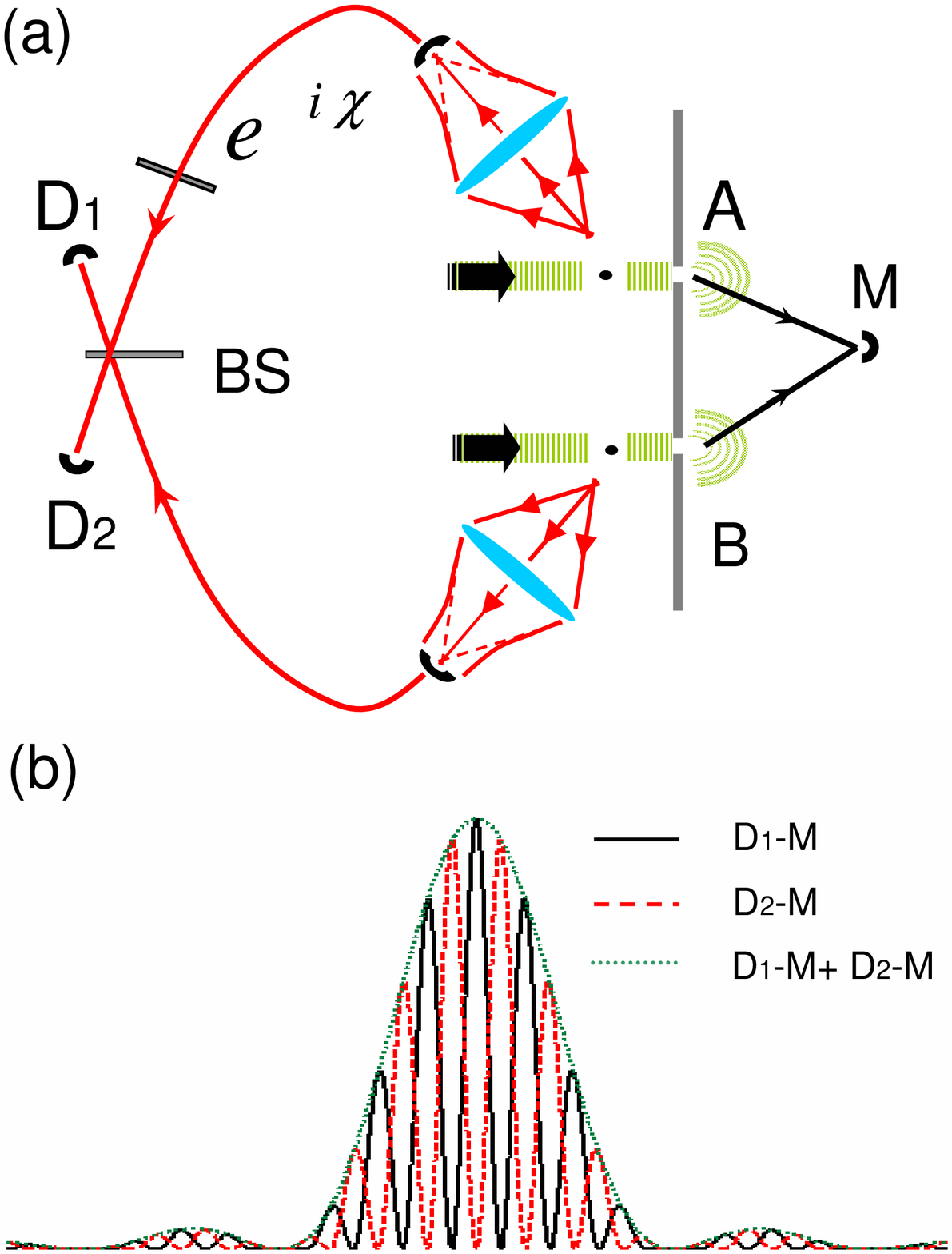}
\caption{(color online) a) Quantum eraser experiment using two
microscopes for recording a photon spontaneously emitted by the
atom close to the aperture A or B.  The photon is oriented through
optical fibers and a 50-50 beam splitter (BS) before reaching one
of the two detectors $D_1$ or $D_2$. The coincidence measurement
atom-photon can be used to rebuild molecular fringes. b) Depending
on the phase shift $\chi$ one can observe fringes or anti-fringes.
If the correlation $D_1$-M gives us the fringes pattern
$|\Psi_{+}|^{2}=\frac{1}{2}|\Psi_{A}+\Psi_{B}|^{2}$ (black curve)
then the correlation $D_2$-M gives
$|\Psi_{-}|^{2}=\frac{1}{2}|\Psi_{A}-\Psi_{B}|^{2}$ (red dashed
curve). The sum of both patterns give us the initial distribution
$|\Psi_{A}|^{2}+|\Psi_{B}|^{2}$ (green dotted envelope).}
\label{fig2}
\end{figure}
 From semi-classical physics this behavior is
prohibited \cite{Scully,RemarksScully}. The most famous quantum
eraser proposal was given by Scully \emph{et al.} in \cite{Scully}
and is based on the micromaser two-cavities setup described in
section
II. \\
Less know is however the fact that quantum eraser could be in
principle realized with the ``Heisenberg'' setup  described in
section II and based on spontaneously emitted photons
\cite{Pfau2}. The principle of this proposal is sketched on
Fig.~2a. Detection of single photons emitted in the close vicinity
of A and B is ensured by microscope objective(s) able to resolve
clearly the apertures region. The photon state coming from A is
oriented through an optical fiber to the first entrance of a 50-50
beam splitter (BS). The second photon state follows a similar path
to the second entrance of the beam splitter. The recombination of
the photon states at the exit outports  $D_1$ and  $D_2$ erases
the which path information since we have no way to know where the
photon was coming from. One could thus retrieve fringes or
anti-fringes by changing the phase $\chi$ between the photon paths
and by detecting in coincidence the molecule arrival on a screen
in $M$ (this experiment is very close to the one realized in
\cite{Kim}). If the Heisenberg experiment described in section II
(Eqs.~4-12) was effectively based on (semi) classical momentum
kicks associated with the emission of a particule the quantum
eraser discussed here could not work! This shows that the semi
classical picture is not adapted even in ``canonical'' experiments
like those described by
Einstein \cite{Bohr1} of Feynman \cite{Feynman}.\\
However, the important word in any quantum eraser experiment is
\emph{coincidence} (i.~e., correlation photon-atom). Indeed, if it
would be possible to erase the fringes without correlating the
detections of the photon and atom then it could be possible to
realize faster than light communication \cite{Jaynes}. The
argument against this possibility is the same as the one used by
Bell in the context of Einstein Podolsky Rosen experiments
\cite{Bell}. If we do not compare the data coming from the two
exits $D_{1,2}$ with the molecule arrival we must find the initial
molecule profile without fringes (see Fig.~2b). This is
particulary clear from the fact that we have \cite{Scully}
\begin{equation}
|\Psi_{+}(\mathbf{r},t)|^{2}+|\Psi_{-}(\mathbf{r},t)|^{2}=|\Psi_{A}(\mathbf{r},t)|^{2}+|\Psi_{B}(\mathbf{r},t)|^{2}.
\end{equation}
It must be remarked that this rebuttal of faster than light
communication gives us a solution for the paradox concerning the
coexistence of momentum transfer with quantum erasure. Indeed,
Eq.~29 is tantamount to argumentations based on decoherence theory
which say that the entanglement with the environment is
responsible for the loss of coherence observed during a which path
experiment (we must sum the two quantum eraser patterns
$|\Psi_{+}(\mathbf{r},t)|^{2}$ and $|\Psi_{-}(\mathbf{r},t)|^{2}$
in order to obtain the complete pattern given by
$|\Psi_{A}(\mathbf{r},t)|^{2}+|\Psi_{B}(\mathbf{r},t)|^{2}$, e.~g., Eq.~ 12 or 15).\\
Now we should remember that the discussion of momentum transfer in
section II tells us that we have necessarily to sum all the
individual patterns associated with different $|\mathbf{k}\rangle$
in order to obtain the full diffraction pattern without fringes.
The present discussion is thus identical to the previous one but
is done in the base $|\gamma_{\pm}\rangle$ instead of
$|\mathbf{k}\rangle$. This is completely consistent with section
IIIA discussing the observer freedom on the choice of the detector
basis $|\xi\rangle$.\\
In conclusion, since all the argumentation presented in section II
is only based on decoherence theory it means that reasonings based
on quantum erasers are rigorously equivalent to those based on
momentum transfer presented in this article. The quantum eraser
corresponds thus precisely to procedures of projection for
measuring the momentum transfer. This is visible from Eqs.~23-25
which show indeed that the phase shift between the two quantum
eraser patterns $|\Psi_{\pm}(\mathbf{r},t)|^{2}$ corresponds to
momentum
transfers of $p_{+}=0$ and $p_{-}=h/2d$ respectively (e.~g, to phase shifts of $\phi_{-}=0$ and $\phi_{+}=\pi$ respectively).\\
It is well known that with a lens focussed on the two-apertures
plane one can in principle realize a which path experiment. The
experimentalist has the choice to position the photon detectors at
any distance behind the lens. If the detectors are located in the
image plane of the lens one can distinguish the path followed by
the atom in the interferometer. However, if the photon detectors
are positioned in the back focal plane of the lens (i.~e, Fourier
plane) then the observer can record the photon momentum
distribution in the base $|\mathbf{k}\rangle$. As discussed in
section II (i.~e., Eq.~6) the joint distribution of probability
photon-atom reveals atom fringes with a phase shift $k_{x}d$ which
corresponds to a transverse momentum transfer
$p_{\mathbf{k}}=\hbar k_{x}$. From this discussion it is thus in
principle possible to realize a quantum eraser experiment by
correlating photon detection in the back focal plane of the lens
with the detection of an atom on the screen. This shows once again
that the quantum eraser is indeed connected to the definition of
momentum transfer used in this article. The same procedure using a
lens could be (in principle) realized with the setup described in
\cite{Scully}. Such an experiment would correspond to a measure of
the momentum distribution $p_{\mathbf{k}}=\hbar k_{x}L/d$ (see
Eqs.~14-16). However, the presence of the cavities walls
constraints in practice such observations with a
lens~\cite{lastremark}. To realize a quantum eraser it is thus
easier to consider the basis $|\gamma_{\pm}\rangle $ instead of
the
basis $|\mathbf{k}\rangle$.\\
The apparent contradiction between the existence of quantum
erasers on one hand and the existence of momentum transfer on the
other hand results from the (implicit) use of a classical
definition of momentum. Once again, with such definition the
quantum eraser would not be possible. Here our definition of
momentum transfer, e.~g.~,
\begin{equation}
p_{\xi}=\hbar\phi_{\xi}/d\end{equation} does not have this
handicap. This definition is indeed purely quantum and represents
the most direct generalization of the usual definition used for
the case of Heisenberg's like experiment in Eqs.~4-12. It can be
applied to any double-apertures which-path or quantum eraser
experiments (like the ones described in \cite{Scully}).
\section{Summary}
 Our discussion of wave-particle duality and of momentum
transfer is directly based on the very basis of quantum mechanics
which involves only observable quantities associated with states
$|\xi\rangle$. This means that the momentum transfer is always in
principle experimentally accessible to the observer. This clearly
shows the pertinence of our definition in a physical discussion
concerning the meaning of momentum transfer. Our analysis show
additionally that momentum transfer and quantum eraser are not two
independent concepts and that they are in fact two formulations of
the same decoherence argumentation applied two double-aperture
experiments. Clearly, most of the formalism used here was already
known \cite{Tan}. However, the analysis in term of momentum of
experiments like the one proposed in \cite{Scully} has to our
knowledge never been done, consequently missing some fundamental
subtleties of the interpretation.


\begin{references}
\bibitem{RemarksScully}
Consider a model in which the atom is treated by quantum mechanics
but where the photon is treated classically as a point particle.
Since after its emission the photon will follow a given trajectory
we conclude that what ever we could do by acting locally on the
photon will not change the fact that atom fringes were already
destroyed by the emission process.
\bibitem{lastremark}
This handicap could be in principle overcome by considering a
hypothetical particle crossing the two cavities (in the
transversal direction $x$) without interacting with the walls. If
such particle absorbs the photon one could compare its momentum
before and after the interaction and thus rebuild the atom
interference pattern (e.~g., realize a quantum eraser in the basis
$|\mathbf{k}\rangle$) by correlating the information obtained with
the recording of the atoms.
\end{references}
\end{document}